# Strain-induced suppression of thermochromism in divalent cobalt molybdate thin films

Kiri Van Koughnet[1], Joey Williamson[1], Kane Hill[2,3], Robert G. Buckley[1,4], Yurii G. Pashkevich[5,6], Sergej M. Orel[6], Eric Lebraud[7], Manuel Gaudon[7], Peter P. Murmu[8], Sarah Spenser[1], Dominique Appadoo[9], Shen V. Chong[1,4], Fryderyk Lyzwa[2,3,4]

[1]*Robinson Research Institute, Victoria University of Wellington, P.O. Box 33—436, Lower Hutt 5046, New Zealand*
[2]*Department of Physics, Photon Factory, University of Auckland, 38 Princes Street, Auckland 1010, New Zealand*
[3]*Te Whai Ao Dodd-Walls Centre for Photonic and Quantum Technologies, New Zealand*
[4]*MacDiarmid Institute for Advanced Materials and Nanotechnology, P.O. Box 600, Wellington 6140, New Zealand*
[5]*Department of Physics and Fribourg Center for Nanomaterials, Chemin du Musée 3, University of Fribourg, Fribourg, Switzerland*
[6]*O.O. Galkin Donetsk Institute for Physics and Engineering NAS of Ukraine, 03028 Kyiv, Ukraine*
[7]*Univ. Bordeaux, CNRS, Bordeaux INP, ICMCB, UMR5026, 87 avenue du Dr. A. Schweitzer, Pessac, F-33608, France*
[8]*National Isotope Centre, Earth Sciences New Zealand, 30 Gracefield Rd, PO Box 30368, Lower Hutt 5010, New Zealand*
[9]*THz/Far-IR Beamline, Australian Synchrotron, 800 Blackburn Rd, Clayton, VIC, 3168 Australia*

## Abstract

Thermochromic oxides provide a platform for coupling lattice, electronic, and magnetic degrees of freedom, with divalent cobalt molybdate ($CoMoO_4$) as a prototypical example. Despite extensive studies on powders, its thin-film behaviour - critical for device applications - remains largely unexplored. Here, we present a combined experimental and theoretical investigation of $CoMoO_4$ thin films, including the first terahertz-to-visible/UV optical spectra of β-$CoMoO_4$ in thin-film form. In contrast to bulk powders, which undergo a first-order β→$\alpha$ structural transition near 230 K, the thin films retain the high-temperature β-phase across the entire temperature range. We show that microstrain fundamentally reshapes the phase landscape, suppressing thermochromism and stabilizing the high-temperature phase. The optical spectra reveal pronounced anomalies in the lattice and electronic responses, including a softening of a low-energy, cation-dominated phonon (42 $cm^{-1}$) upon cooling, in contrast to conventional hardening of higher-energy modes. This behaviour indicates incipient atomic displacements analogous to those driving the bulk β→$\alpha$ transition, despite the absence of a structural phase change, and saturates between 200 and 150 K. Temperature-dependent X-ray diffraction confirms persistent β-phase symmetry with increasing microstrain, consistent with strain-induced frustration of the first-order transition. At higher energies (0.12–3.7 eV), the optical response exhibits a blue-shift of $Co^{2+}$crystal-field transitions and charge-transfer excitations, indicating a strain-enhanced ligand field. Supported by structural refinements and theoretical calculations, we identify crystal field strengthening as the key mechanism stabilizing the β-phase. These results establish strain as a thermodynamic lever to control phase stability and functional properties in thermochromic oxides.

## Introduction

Allotropic phases of the divalent molybdates display contrastingly changing colors while exhibiting different magnetic states [1-4]. This range of interesting behaviours has been exploited in numerous applications, including catalysis, batteries, and sensing [5-7]. Of particular interest is the prominent color change exhibited by the high-pressure and low-temperature divalent molybdate $\alpha$-$CoMoO_4$ phase, in changing from green to the structurally different purple β-$CoMoO_4$ high-temperature and low-pressure phase; both phases of $CoMoO_4$ are stable at ambient conditions [1,3]. For powder $CoMoO_4$, the transition temperatures between these two phases have been measured through optical, XRD and Raman measurements, revealing a hysteretic phase transition: $T(\beta \rightarrow \alpha)$ at 230 K upon cooling, and $T(\alpha \rightarrow \beta)$ at 780 K upon heating [1-3]. Both phases crystallise in the monoclinic C2/m space group and have $Co^{2+}$ ions in a distorted octahedral site ($CoO_6$) [1-3,8]. The two phases differ in the coordination of their $Mo^{6+}$ ions: β-$CoMoO_4$ has a structure in which the $Mo^{6+}$ cations form isolated distorted tetrahedra ($MoO_4$), whereas $\alpha$-$CoMoO_4$ has a structure where the $Mo^{6+}$ ions form $Mo_4O_{16}$ tetramers via edge sharing, leaving the $Mo^{6+}$ ion in a distorted octahedral site ($MoO_6$) [1-3,8]. Figure 1 displays the edge sharing $Mo_4O_{16}$ tetramers and isolated $MoO_4$ tetrahedra for $\alpha$-$CoMoO_4$ and β-$CoMoO_4$, respectively.

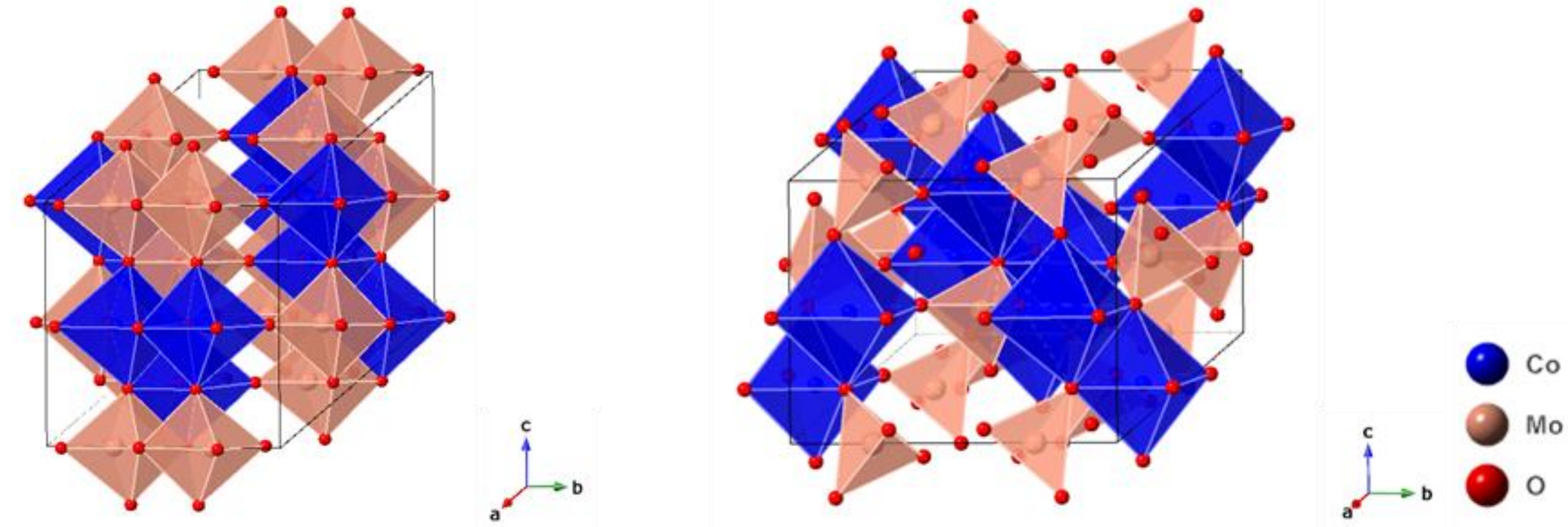

**Figure 1. Crystal structure of $CoMoO_4$**. Images are generated by CrystalMaker, illustrating the edge sharing $Mo_4O_{16}$ tetramers in the α-phase (left) and the isolated $MoO_6$ tetrahedra in the β-phase (right).

Most of the work reported on $CoMoO_4$ was undertaken on powder samples [1,3,8,9]. Studying single crystals, instead of powders, would be ideal for investigating the electronic and magnetic properties of $CoMoO_4$ arising from anisotropy. However, the potential for cracks and fracturing during temperature-dependent studies undergoing the thermochromic first-order phase transition of this molybdate between $\alpha$ to β phases may introduce signal artefacts. This is because upon cooling ($\beta \rightarrow \alpha$) or heating ($\alpha \rightarrow \beta$), $CoMoO_4$ undergoes a 6% change in the unit cell volume [3], which as observed in single crystal $CuMoO_4$ [10] and $CuMoO_4$ crystallites [11], may lead to cracking of the crystals. Another issue is that large millimetre-sized $CoMoO_4$ single crystals are not readily available [12-14]. For many applications however, thin films would be the materials form of choice. However, there have only been a few studies based on $CoMoO_4$ thin films in the literature [15-17] including our recent report on the magnetic properties of mixed $\alpha$- and β-phase $CoMoO_4$ thin films [18].

Here, we report a combined structural, magnetic and optical study of thin films of one phase, namely β-$CoMoO_4$, deposited on silicon and sapphire substrates. Interestingly, we find that the β-phase is stable in thin film form down to the lowest temperatures ($T$ = 12 K) confirmed by X-ray diffraction (XRD). To enable the utilization of β-$CoMoO_4$ for applications, we are exploring the low-energy excitations (infrared-active phonons) and electronic transitions in pure β-$CoMoO_4$ thin films, and compared the finding with the properties of powder samples. To this end, and because of the enhanced stability of thin films of the β-phase, we report the first optical spectra of β-$CoMoO_4$ in thin film form across a wide spectral range from the terahertz to the visible/UV. We studied the temperature dependence of the phonons between 300 K and 150 K and observed an anomalous softening of the lowest energy (42 $cm^{-1}$) cation-dominated phonon. In the visible/UV spectral range we observe the 300 K behaviour across spectral regions, probing the $Co^{2+}$ *d-d* band transitions and the oxygen *2p-to-Mo-d*-orbital inter-band or charge transfer transition.

## Experimental Methods

### Film Deposition

β-$CoMoO_4$ films were deposited at room temperature using magneto-sputtering on a variety of substrates. The XRD and magnetic studies were carried out on films deposited on silicon substrates that had been thermally treated resulting in a 200-300 nm $SiO_x$ surface layer. For the optical studies, films were deposited on double-side polished silicon and sapphire substrates. The 1-inch $CoMoO_4$ sputtering target, with a purity of 99.9%, was supplied by ACI Alloys, Inc. (USA). In all cases, the films were grown using RF magnetron sputtering with a power of 30 Watts at room temperature, under a partial pressure atmosphere of 6 mTorr of Argon, and post-deposition annealed in-situ at 700 °C for 5 hours in 1 mTorr of oxygen. For a target film thickness of 200 nm, the film was deposited for 1 hour, although to improve our sensitivity for the optical study, a 2-hour deposition duration was employed to achieve film thicknesses of 450 nm. The film thickness was determined by scanning electron microscopy employing a Hitachi SU7000 system, where it was found that a target film thickness of 200 nm resulted in a measured thickness of 185 ± 15 nm, and that of the thicker film in a thickness of 470 ± 20 nm. Herein, we will refer the films in their nominal thicknesses of 200 nm and 450 nm, unless otherwise specified.

### Structural and compositional characterisation

X-ray diffractograms were acquired on a Rigaku Smartlab system using Co $K_\alpha$ X-ray radiation at a wavelength of 1.7889 Å at a grazing angle of incidence (2°) to suppress any contribution from the substrate. Temperature-dependent XRD patterns were collected between room temperature and 12 K on a PANalytical X'Pert Pro powder diffractometer equipped with an Oxford Instruments liquid helium cryostat (6K) employing copper radiation in the usual $\theta$-$2\theta$ configuration.

The compositional profile of the β-$CoMoO_4$ film was determined using Rutherford backscattering spectrometry (RBS). RBS was performed with a 2.0 MeV $^4He^+$ probe beam, a beam diameter of 1 mm, and a surface barrier detector positioned at a 165° backscattering angle to measure the energy and yield of backscattered particles. With this probe energy, the depth profile will be in the μm range; therefore, RBS provides the average composition of the entire 200 nm thick cobalt molybdate film. The film's composition was determined using the Rutherford Universal Manipulation Program (RUMP) computer simulations [19].

X-ray photoelectron spectroscopy (XPS) measurements were conducted on an ESCALAB250Xi Thermo Scientific XPS system using monochromatic Al Kα radiation (1486.6 eV) set at 120 W. The measurements were undertaken with perpendicular incident X-rays and a photoelectron take-off angle of 90°, enabling a depth profile of ≤ 50 Å. XPS scans were taken with 0.1 eV steps and with a pass energy of 20 eV. The binding energy was referenced to the XPS C 1s peak of adventitious carbon at 284.4 eV. The quantification of atomic compositions was obtained from the XPS peak areas using a Shirley background function with built-in cross-sections adjusted.

### Magnetic characterisation

Magnetization data on the $CoMoO_4$ thin films were acquired using a Quantum Design MPMS-7 SQUID magnetometer with the applied magnetic field parallel to the surface of the films and taking measurements between 2.2 K and 300 K in magnetic fields up to 6 T. Note, the magnetisation curves presented have not been background subtracted.

### Optical measurements

We conducted a comprehensive optical characterization of the $CoMoO_4$ films, ranging from Terahertz (THz) to the Ultraviolet (UV)spectral range. Across this wide spectral range, we employed two substrates: silicon in the THz -mid-infrared, and sapphire in the near-infrared/visible/UV.

The THz/Far-infrared transmission experiments were performed at the Australian Synchrotron's Terahertz beamline. Transmission spectra were measured between 20 and 685 $cm^{-1}$ (2.5 - 85 meV) at a resolution of 1 $cm^{-1}$ using a Bruker IFS125 spectrometer equipped with a 6 $\mu$m Mylar beamsplitter and a Si bolometer. The samples were mounted on the translation stage of a cryostat optically coupled to the IFS125 spectrometer and studied from room temperature down to 150 K. These spectra were combined with 300 K infrared (IR) to ultraviolet (UV) spectra, ranging from 150 $cm^{-1}$ (18 meV) to 30,000 $cm^{-1}$ (3.7 eV) and taken at a laboratory-based Bruker Vertex 80v interferometer, measuring both the reflectivity and transmission. The technique has been reported elsewhere [20,21]. Further information on the signal intensity can be found in the supplementary materials in Figure S1. The spectral transmission was determined by taking the ratio of the transmitted and incident signal intensities, while the reflectivity was determined relative to an evaporated Al film and corrected by employing literature values for the aluminium reflectivity [22]. The reflectivity was measured at near-normal incidence to the sample surface plane, to study the film's in-plane response. The uncertainty in the magnitude of these quantities is ±2 %, based on the reproducibility of the spectra.

The reflectivity and transmission spectra of the film-substrate stack were simultaneously modelled, by means of the public domain software package RefFit [23]. By using a Kramers-Kronig consistent sum of Drude-Lorentzian functions, we simulate the energy-dependent dielectric function, $\varepsilon(\omega) = \varepsilon_1(\omega) + i\varepsilon_2(\omega)$, where:

$$\varepsilon(\omega) = \varepsilon_\infty + \sum_k^N \frac{S_k^2}{\omega_{0k}^2 - \omega^2 - i\omega\gamma_k}$$

$\varepsilon_\infty$ represents the high-frequency dielectric constant, $\omega$ is the wavenumber, $\omega_0$ is the oscillator frequency, $S$ is the oscillator strength, $\gamma$ is the damping coefficient for each oscillator $k$. The optical response of the substrate was incorporated after taking independent measurements of $R$ and $T$ of a bare substrate, i.e. of either silicon or sapphire, and then deriving a model substrate dielectric function. The film thickness, as determined above, was used as a fixed parameter.

## Results and Discussion

### Sample growth, Structure, and Chemical Composition

Figure 2(a) exhibits an XRD pattern collected at 2° X-ray incident angle at ambient temperature for a 200 nm film grown on a Si substrate. The XRD pattern confirms the exclusive presence of β-phase $CoMoO_4$, showing all the reflections that are expected for a monoclinic crystal structure with space group (C2/m) [24]. Notably, there are no additional XRD reflections from impurities, nor Bragg peaks at 16.4° and 33.1° 2$\theta$ for (110) and (220) Miller indices, respectively, which would indicate the presence of $\alpha$-$CoMoO_4$. Furthermore, like the powder XRD patterns found in previous studies [1,2], there is a preferential growth plane (002) at 30.7° 2$\theta$ as indicated in Figure 2(a). The 300 K lattice parameters of a monoclinic β-phase $CoMoO_4$ film on silicon are listed in Figure 2(b) and are in good agreement with the values found previously, for films on both silicon and sapphire, and for powder samples [18]. A few films did however exhibit a minor (4 %) fraction of $\alpha$-$CoMoO_4$, which we will discuss further below.

In addition, temperature-dependent XRD patterns were collected between 300 K and 12 K for two different thin films deposited on silicon substrates (Figure S2 and S3). These demonstrate that, unlike for powder samples, the β-phase thin film β-$CoMoO_4$ is stable down to 12 K and that the samples do not exhibit a thermochromic phase transition at ~230 K. The reflections attributed to the small fraction of $\alpha$-$CoMoO_4$ observed in one of the films (~4 %, as noted above and detailed in Figure S3) also remain unchanged in both position and intensity down to 12 K. While we observed the expected contraction of the β-phase unit cell volume with decreasing temperature, the contraction is not uniform, saturating at 150 K with no measurable change between 150 K and 12 K (Table S1). The measured changes in the *a*-axis and the *c*-axis were not significant, but there was an observable contraction between 300 K and 150 K of the *b*-axis of 0.1% and a 0.9% decrease in the monoclinic angle (β). The same lack of temperature dependence was exhibited by the minor $\alpha$-$CoMoO_4$ phase.

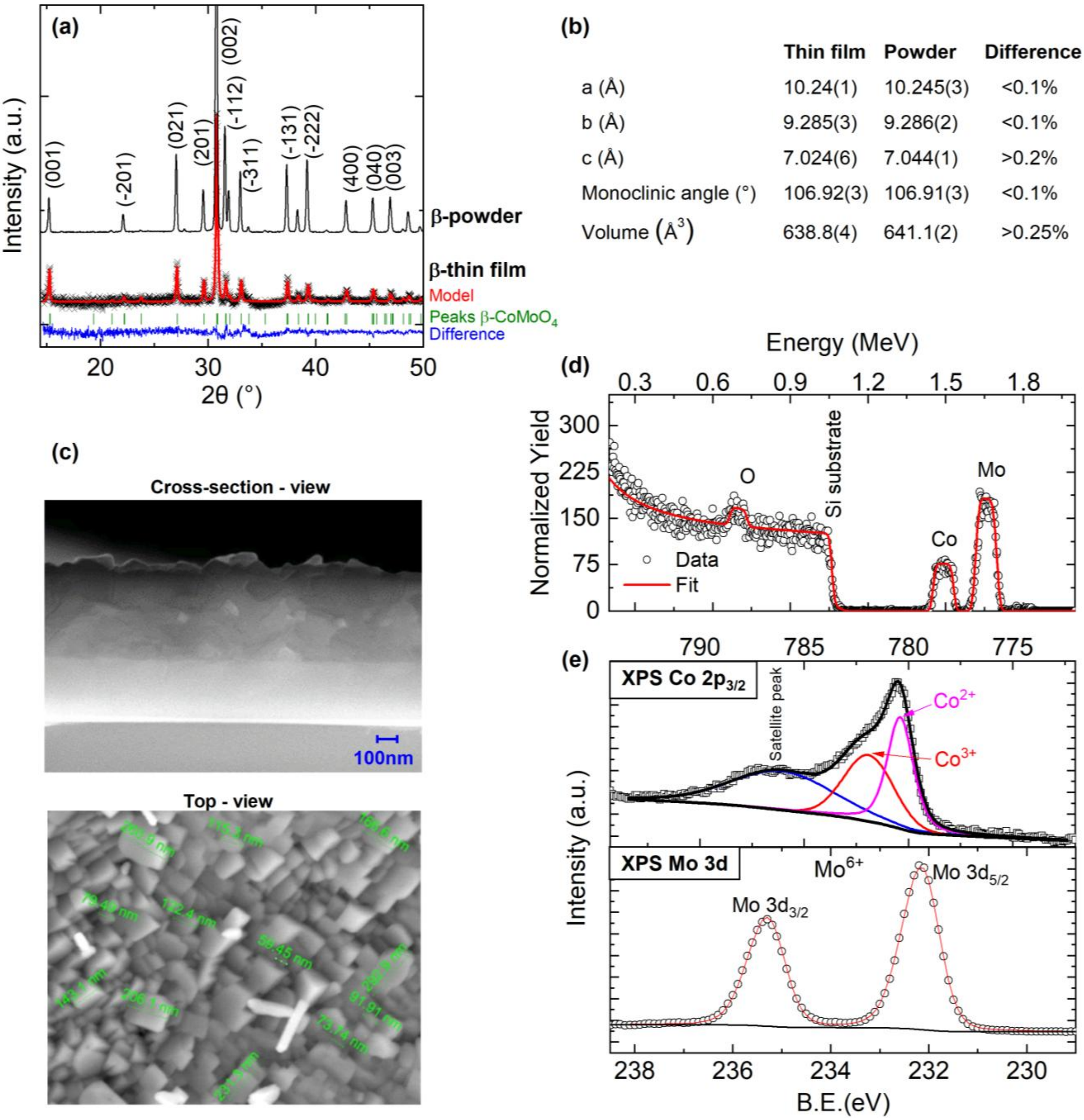

| | Thin film | Powder | Difference |
|---|---|---|---|
| a (Å) | 10.24(1) | 10.245(3) | <0.1% |
| b (Å) | 9.285(3) | 9.286(2) | <0.1% |
| c (Å) | 7.024(6) | 7.044(1) | >0.2% |
| Monoclinic angle (°) | 106.92(3) | 106.91(3) | <0.1% |
| Volume (Å$^3$) | 638.8(4) | 641.1(2) | >0.25% |

**Figure 2. Growth, Structure and Chemical Composition of β-phase $CoMoO_4$ thin films. (a)** Experimental XRD pattern for the 200 nm-thick film grown on Si, and β-phase $CoMoO_4$ powder plotted for comparison. The red-colored trace shows the calculated XRD pattern for the film obtained from structural refinement using GSAS, and the resulting lattice parameters are listed in **(b)**. **(c)** SEM cross-sectional view and top view of a 450 nm-thick β-$CoMoO_4$ film showing the formation of densely packed sub-micron crystallites. A thicker film was employed for ease of imaging. **(d)** Rutherford backscattering spectrum, **(e)** XPS Co $2p_{3/2}$ and Mo 3d core level of a 200 nm-thick β-$CoMoO_4$ thin films for identification of the chemical composition and oxidation states.

To elucidate the absence of thermochromism in our films, we undertook an SEM study of the film morphology. Figure 2(c) shows a typical SEM micrograph illustrating the formation of densely packed sub-micron crystallites in our films that range from 60 to 320 nm. Studies on $CoMoO_4$ [5,25] and $CuMoO_4$ [25] have shown that low-dimensional structures tend to stabilise low-coordination structures, namely the tetrahedral $MoO_4$ in $CoMoO_4$ and $CuMoO_4$, and the square-pyramidal $CuO_5$ in the high-temperature $CuMoO_4$ phase. As the particle size decreases, the ability of these low-coordination structures to transform into the respective octahedral coordination ($MO_6$) at low temperatures becomes increasingly challenging.

The dominant role of strain in suppressing the phase transition in our $CoMoO_4$ films, structural refinements (GSAS-II [26]) is illustrated by the temperature-dependent XRD patterns (Table S1). This allowed us to estimate a value for the microstrain [27]. The instrumental broadening was calculated from measurements of a NIST SRM1976 $Al_2O_3$ standard reference. The particle size parameter was fixed at 4 μm for all refinements such that non-instrumental broadening effects were modelled solely by the microstrain parameter. While the resulting absolute values for the microstrain are not representative of the actual strain occurring in the sample, the temperature-dependent changes well-reflect the actual changes in the strain. Following this approach, the estimated microstrain value increased by 3.7% as the samples were cooled to 150 K and saturating at lower temperature, resulting in a total 4.5% strain-change between 300 K and 12 K. This demonstrates that microstrain in the films increases with decreasing temperature across the expected thermochromic transition temperature, despite the absence of the transition in these thin films. Analysis of 300 K XRD measurements on β-phase $CoMoO_4$ powder, exhibiting a thermochromic phase transition at ~230 K, also show that microstrain is 2-3 times larger in thin films than in powders of β-$CoMoO_4$, providing further evidence for enhanced strain state in the films. While the factors that govern the irreversibility of the β → α transition remains an open question [1], our results indicate that strain plays an important role, consistent with the arguments of Ref. [3].

Further composition analyses of bulk and of the surface of the thin films were carried out using Rutherford backscattering spectrometry (RBS) and X-ray photoelectron spectroscopy (XPS), respectively. Figure 2(d) shows that the best-fit RUMP-simulated RBS spectrum was obtained using a composition ratio of Co:Mo:O = 1:1:4, consistent with the film's target stoichiometry as $CoMoO_4$. XPS probes the surface electronic states of the $CoMoO_4$ films, and Fig. 2(e) shows the XPS Co 2p spectrum which exhibits two Co $2p_{3/2}$ & Co $2p_{1/2}$ peaks at 780.6 eV & 796.6 eV, respectively. Our Co $2p_{3/2}$ peak was modelled via three components rather than two, in agreement to Ref. [12,28]. We identified two cobalt species at binding energies of approximately 780.5 ± 0.2 eV and 782.5 ± 0.2 eV, along with a shake-up satellite at 787.1 ± 0.3 eV. The lower binding energy peak is assigned to $Co^{2+}$, supported by an observed spin-orbit splitting (SOS) of 16.0 eV, while the higher energy peak corresponds to $Co^{3+}$, which typically exhibits a slightly lower SOS of around 15.5 eV [29] (Figure S4). The $Co^{3+}$ species is low-spin (S = 0, diamagnetic) and thus contributes negligibly to the magnetic signal (see Figure 3), whereas the $Co^{2+}$ is in the high-spin state (S = 3/2). The ratio of $Co^{2+}/Co^{3+} \approx 1$ suggests an absence of the $Co_3O_4$ species (with $Co^{2+}/Co^{3+} \approx 0.5$ [30]) on our thin film surfaces.

The two additional peaks centred at 232.2eV (235.3 eV) are attributed to Mo $3d_{5/2}$ (Mo $3d_{3/2}$,), as shown in the bottom panel of Figure 2(e). The spin-orbit-splitting between these two peaks is 3.1 eV, and only one species could be fitted to this peak with a FWHM ~0.83 eV that is assigned to the fully oxidised $Mo^{6+}$ oxidation state. No occurrence of the $Mo^{4+}$ could be observed in any of the thin films. The lattice oxygen ($O^{2}$) concentration (not shown) to that of molybdenum (O/Mo atomic ratio) is 4.3±0.2, which is consistent with a $MoO_4^{2-}$ molybdate anion. This and the results from grazing angle XRD confirm the absence of $Co_2Mo_3O_8$ (with $Mo^{4+}$ species), which tends to form in a low-oxygen partial pressure environment. The excess oxygen in the O/Mo ratio can also explain the presence of higher oxidised $Co^{3+}$ species observed in XPS Co 2p spectrum. We also note that the line shape of the shake-up satellite in the Co $2p_{3/2}$ trace has a characteristic flat peak with a steep edge on the high binding energy side. This distinct shape is characteristic of $CoMoO_4$, as $Co_3O_4$ and CoO have different line shapes [5,25]. Hence, the XPS results confirm the formation of a slightly oxidised stoichiometric $CoMoO_4$ thin film.

## Magnetism

Figure 3 displays the field-cooled (FC) magnetization behaviour, taken at 50 mT. The magnetisation *M* increases with decreasing temperatures, rising to a peak at approximately 10 K, which is due to an onset of antiferromagnetic order [18]. The inset shows the field loops acquired below and above the magnetic ordering temperature of 10 K. The field loops below 10 K show a single sharp magnetic transition at ± 2.2 Tesla without hysteresis in the low field regions (Figure S5). The absence of hysteresis in these thin film samples is a different result from that reported in our previous work from powder samples [18]. This may be due to a lower defect content (e.g., oxygen defects) in the present films that could act as pinning centres, the absence of an $\alpha$-$CoMoO_4$ fraction, and/or a more oxidised $CoMoO_4$ surface (less high-spin $Co^{2+}$ and the presence of low-spin $Co^{3+}$) in these films as shown by XPS analysis. Field-dependent magnetisation measurements on Mn- and Ni- molybdate single crystals revealed two spin-flop transitions with the applied magnetic field (*H*) being parallel to the easy-axis (D), and only one transition when $H \perp D$ [31,32]. Polycrystalline $CuMoO_4$ demonstrated the existence of two spin transitions [33].

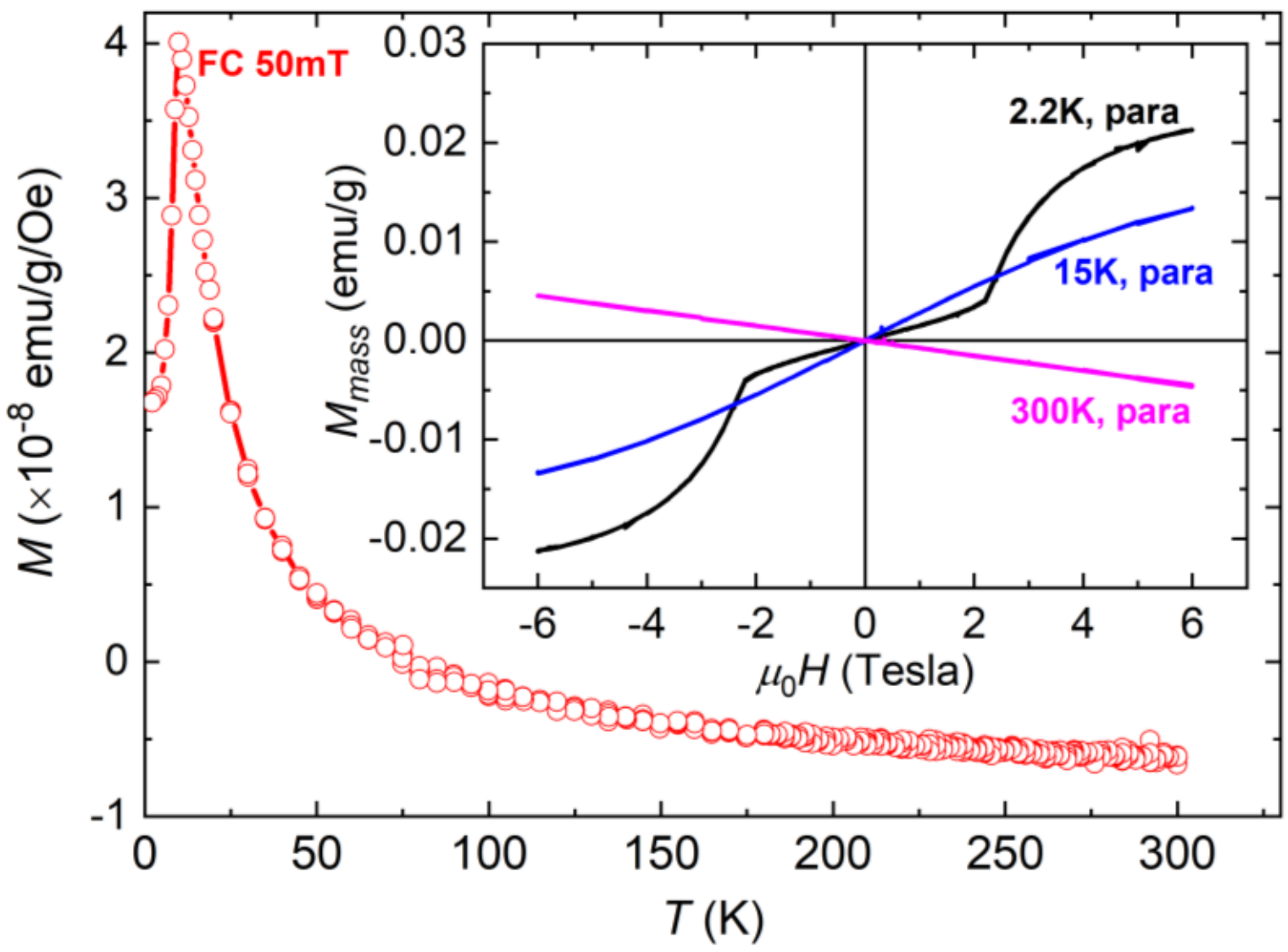

**Figure 3. Magnetic behaviour.** Temperature-dependent magnetisation of a $CoMoO_4$ film exhibiting an antiferromagnetic transition (AFM) at ~10 K. Inset: Field-loops up to +/- 6 T, with the applied field being parallel (para) to the surface of the film and at different temperatures, are not exhibiting a hysteresis below the AFM temperature.

Unlike these other molybdates, the field-dependent spin transition observed in our β-$CoMoO_4$ films is sharp and therefore cannot be attributed to spin-flop transitions due to weak magneto-crystalline anisotropy as observed in $MnMoO_4$ and $NiMoO_4$ [34,35] nor breaking of antiferromagnetic exchange terms as observed in $CuMoO_4$ [33]. Instead, the observed spin transition is more reminiscent of a 'spin-flip' transition, which is mediated by strong magneto-crystalline anisotropy and has been observed in other antiferromagnetic materials [36,37]. It is possible that the spin transition in our β-$CoMoO_4$ films is indeed a spin-flip transition, if the anisotropy terms are stronger than the antiferromagnetic ones.

Note, that even though the magnetic structure has been studied for several molybdates in the α-phase, via neutron scattering or Mössbauer spectroscopy, such as for iron molybdate [38], it is so far unexplored for cobalt molybdate in the strain β-phase. Further joint experimental and theoretical studies are needed to better understand the spin states in β-$CoMoO_4$, which can be useful in spin-processing functionalities since we have now successfully fabricated β-$CoMoO_4$ thin films that are stable down to cryogenic temperatures.

## Optical spectroscopy: Synchrotron-based Terahertz combined with bench-top IR-UV spectra

### *Low-energy excitations: Infrared-active phonons*

So far, there exist are only a few studies of the phonon spectrum of β-$CoMoO_4$, as noted in Ref. [34], which is largely due to the technical challenge of applying the necessary pressure, that converts the β-phase to $\alpha$-$CoMoO_4$ in powder samples, on large enough to be used for infrared (IR) spectroscopy. These previous IR reports are limited to room-temperature studies of $\alpha$-$CoMoO_4$ [34,39-41]. We present, to the best of our knowledge, the first IR study of β-$CoMoO_4$ reporting, together with its temperature dependence.

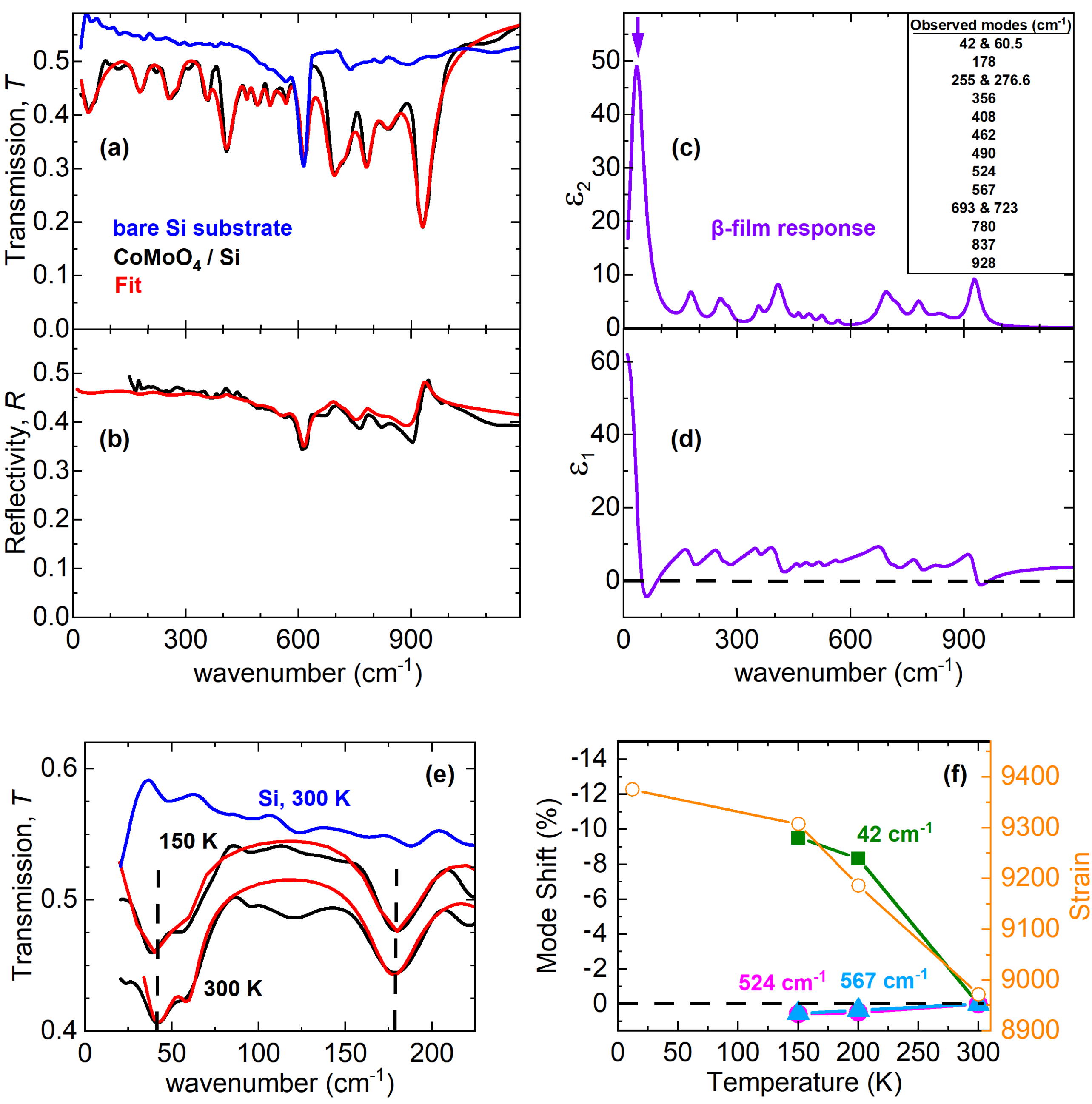

**Figure 4. Low-energy dynamics of a 450 nm-thick β-phase $CoMoO_4$ film.** Room-temperature THz-to-Infrared transmission and reflectivity spectra of the film grown on a silicon substrate (black), of the modelled dielectric function (red) and of a bare silicon substrate (blue) are shown in **(a,b)**. The derived real and imaginary part of the dielectric function ($\varepsilon = \varepsilon_1$,+i*$\varepsilon_2$) of the β-phase $CoMoO_4$ film response are shown in **(c,d)**, illustrating the high strength of the phonon at ~40 cm$^{-1}$. The inset in (c) lists 16 infrared-active modes observed at 300 K. We find three absorption bands that are best fit by two closely spaced Lorentzian functions and displayed as paired modes. The low-energy behaviour at cryogenic temperatures is shown in **(e)**. The dashed lines highlight the 42 cm$^{-1}$ mode that exhibits a softening with decreasing temperature, and a mode at 180 cm$^{-1}$ that exhibits a hardening. **(f)** Percentile mode shift of the low-energy absorption band at 42 cm$^{-1}$ (green, square) compared to the 524 cm$^{-1}$ (magenta, circle) and 567 cm$^{-1}$ (cyan, triangle) bands. Between 300 K and 150 K, all modes above 178 cm$^{-1}$ exhibit a hardening by 0.6 % on average, while only the 42 cm$^{-1}$ shows a softening by 9 %. The observed unit cell strain, calculated from XRD measurements, are plotted for comparison (orange, open circles).

Figures 4(a,b) show the room temperature measured terahertz to infrared transmission (*T*) and reflectivity (*R)* spectra (black curves), across the IR-active phonon region for a 450 nm thick β-$CoMoO_4$ film grown on a silicon substrate. As noted above, the transmission is a combined data set collected on the Australian Synchrotron (20 $cm^{-1}$ to 1200 $cm^{-1}$) and a benchtop-based system (150 $cm^{-1}$ to 1200 $cm^{-1}$). In total we detected 16 absorption bands in the transmission spectrum, which can be assigned to phonons and will be discussed below. The results of simultaneously modelling *T* and *R* (in red) are also displayed in Figures 4(a) and 4(b), respectively. As described above, this spectral region was modelled using a series of Lorentzian oscillators to describe the phonon behaviour of β-$CoMoO_4$. The resulting energy-dependent real and imaginary parts of the dielectric function are plotted in Figures 4(c) and 4(d). Interestingly, the derived dielectric function demonstrates that the low-energy absorption band near 40 $cm^{-1}$ exhibits a high oscillator strength relative to the other modes.

Importantly, with IR-spectroscopy techniques we are generally not sensitive to the $A_g$ and $B_g$ modes, unlike the Raman scattering process. As indicated by the XRD measurements presented above, all the $CoMoO_4$ films studied are in the β-phase. The crystal structure of β-$CoMoO_4$ is complex, with a monoclinic C2/m space group with 8 molecules per unit cell. The $Co^{2+}$ ions are in highly distorted octahedral sites while two pairs of $Mo^{6+}$ cations are in a specific tetragonal coordination in the β-phase, with $C_2$ and $C_s$ site symmetry. Group theory analysis indicates a zone centre representation of

$$\Gamma = 19A_g + 17\ B_g + 14A_u + 19B_u$$

which implies that after subtracting the acoustic modes, there are 33 observable IR active modes, and 36 observable Raman active modes [5,25]. Experimentally, 21 Raman excitations can be seen according to Ref. [5], including one near 65 $cm^{-1}$. Here, we report 16 infrared-active modes including a low-energy mode at 42 $cm^{-1}$; a full list is displayed in the inset of Figure 4(c). Figure 4(e) illustrates the typical temperature dependence (between 300 K and 150 K) of the phonon modes of a 450 nm β-$CoMoO_4$ film from 20 to 225 $cm^{-1}$ wavenumbers (black). These spectra are compared with our model (red) and the ambient transmission spectra of a bare silicon substrate (blue). The spectra measured at 200 K and 150 K are nearly identical, and thus we report only the lower temperature spectra. All modes between 178 $cm^{-1}$ and 685 $cm^{-1}$ exhibit a hardening of about 0.6 % as illustrated by the modes at 524 $cm^{-1}$ and 567 $cm^{-1}$ shown in Figure 4(f), which is consistent with the observed contraction of the unit cell volume (△V/V~0.2 %). In contrast, an anomalous softening of the strong 42 $cm^{-1}$ absorption band is of ~9 %, which is significantly larger in magnitude than the hardening of the high energy modes. We observe a similar softening for the mode at 60.5 $cm^{-1}$. Since the absorption band at 42 $cm^{-1}$ is close to the low-energy end of the reliable spectral range (limit of 20 $cm^{-1}$) due to reproducibility reasons, we measured two nominal film thicknesses of 450 nm and 200 nm and tested several fitting models to best describe the data. For instance, we have separately modelled the absorption at 42 $cm^{-1}$ with one (at 41 $cm^{-1}$) and then two (at 42 and 60.5 $cm^{-1}$) Lorentzian functions; the latter two-oscillator model shows a slightly better fit on the high energy side.

As illustrated in the Supplementary Material (Figure S6), and to ensure repeatability of this anomalous low-energy spectral feature, we have measured the transmission of multiple films (with nominal thicknesses of 200 nm and 450 nm) grown on silicon substrates at ambient conditions and temperatures down to 150 K. The absorption band at 42 $cm^{-1}$ is reproduced in all the measurements on our films and is absent in the bare substrate at any temperature (see Figure 4(e) and Figure S6). In addition, note that Ref. [25] reports a Raman mode at 50 $cm^{-1}$ on a rising background and Costa et al. report a peak near 65 $cm^{-1}$ in $\alpha$-$CoMoO_4$ [1,35]. Although the 42 $cm^{-1}$ band is on the boundary of reliable spectral data range, given the number of films studied and the Raman measurements, the evidence indicates that β-$CoMoO_4$ exhibits a low energy mode at 42 $cm^{-1}$ that softens with decreasing temperature, and is likely dominated by the motion of the Co and Mo ions. Both the appearance of a strong low-energy mode that significantly increases the dielectric constant, and its softening with decreasing temperature, is a well-known behaviour in $SrTiO_3$ (e.g. Ref. [42-44]), which is a quantum paraelectric that exhibits a ferroelectric instability. We find no evidence of a nearby ferroelectric order in our samples, and future experimental and theoretical work is needed. A mode-softening has been previously reported in a temperature-dependent Raman study on $\alpha$- and β-phase $FeMoO_4$ [38]; however, that mode is of much higher energy (~920 $cm^{-1}$), and the percentile softening of <1 % is significantly smaller than in our case (9 %), which suggests that these cases might not be directly comparable.

It is well known that β-$CoMoO_4$ powder undergoes a first-order structural transition at 230 K on cooling to $\alpha$-$CoMoO_4$. Figure 5 illustrates that in β-$CoMoO_4$, the $Mo^{6+}$ ions sit in a distorted tetrahedral site, while in $\alpha$-$CoMoO_4$, the $Mo^{6+}$ ion moves to a distorted octahedral site. This structural transition represents a significant movement of both the $Mo^{6+}$ and several of the coordinated oxygen ions. We note that while it is difficult to separate the impact of grain size and microstrain in our samples from the XRD analysis, we nonetheless propose that the softening of the 42 $cm^{-1}$ mode represents the initiation of the first-order β-$CoMoO_4$ to $\alpha$-$CoMoO_4$ transition, which is suppressed by the strain and grain size in the film on the non-epitaxial silicon substrates employed.

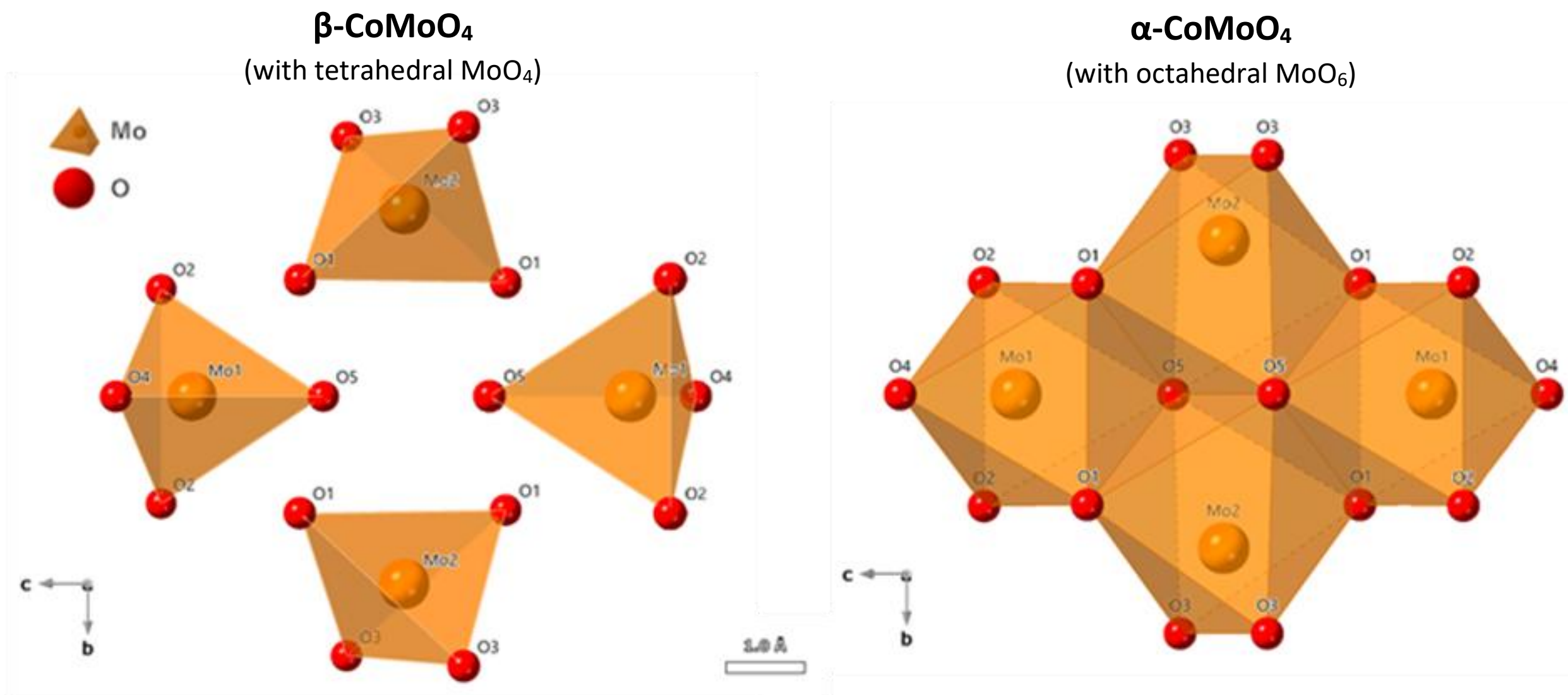

**Figure 5.** Molybdenum ions coordination in β-phase (tetrahedral $MoO_4$) and $\alpha$-phase (octahedral $MoO_6$) $CoMoO_4$.

***Electronic transitions***

Now we turn to the spectra collected from 1,000 $cm^{-1}$ and 30,000 $cm^{-1}$ (0.12 eV to 3.7 eV) and focus on the electronic transitions. Figure 6(a,b) displays the measured reflectivity and transmission (black curves) of a 290 nm thick β-phase $CoMoO_4$ film grown on sapphire. The reflectivity and transmission were simultaneously modelled (red curves) using RefFIT, in analogy to the fitting of the phonon spectral region. In our approach, we first chose the film thickness as a variable to match the interference pattern observed. This resulted in shifts of no more than 10 % from the target thickness. Next, the high-energy dielectric constant was adjusted to match the magnitudes of reflectivity and transmission. Both these adjustments were undertaken in the low-energy non-absorbing region (i.e., for < 4,000 $cm^{-1}$ / ~0.5 eV). To describe the inter-band absorption, we included a large oscillator near the high-energy limit of the data, given the rapid decrease in transmission at around 25,000 $cm^{-1}$. Three absorption bands at lower-energy were then introduced to model the *d-d* band transitions for $Co^{2+}$ using literature energies as the starting values [3,25].

Figure 6(c) displays the energy-dependent real-part of the optical conductivity derived from modelling the reflectivity and transmission, noting that this parameter is directly related to the spectral weight absorbed by the film. A rapid rise in the absorption occurs from about 22,000 $cm^{-1}$ (2.7 eV) that we assign to the fundamental inter-band transition, which in the literature is often referred to as a 'charge transfer band' and this value is in very good agreement with reported values for β-$CoMoO_4$ located in the 2.8-2.9 eV range [1,3,5,25,35].

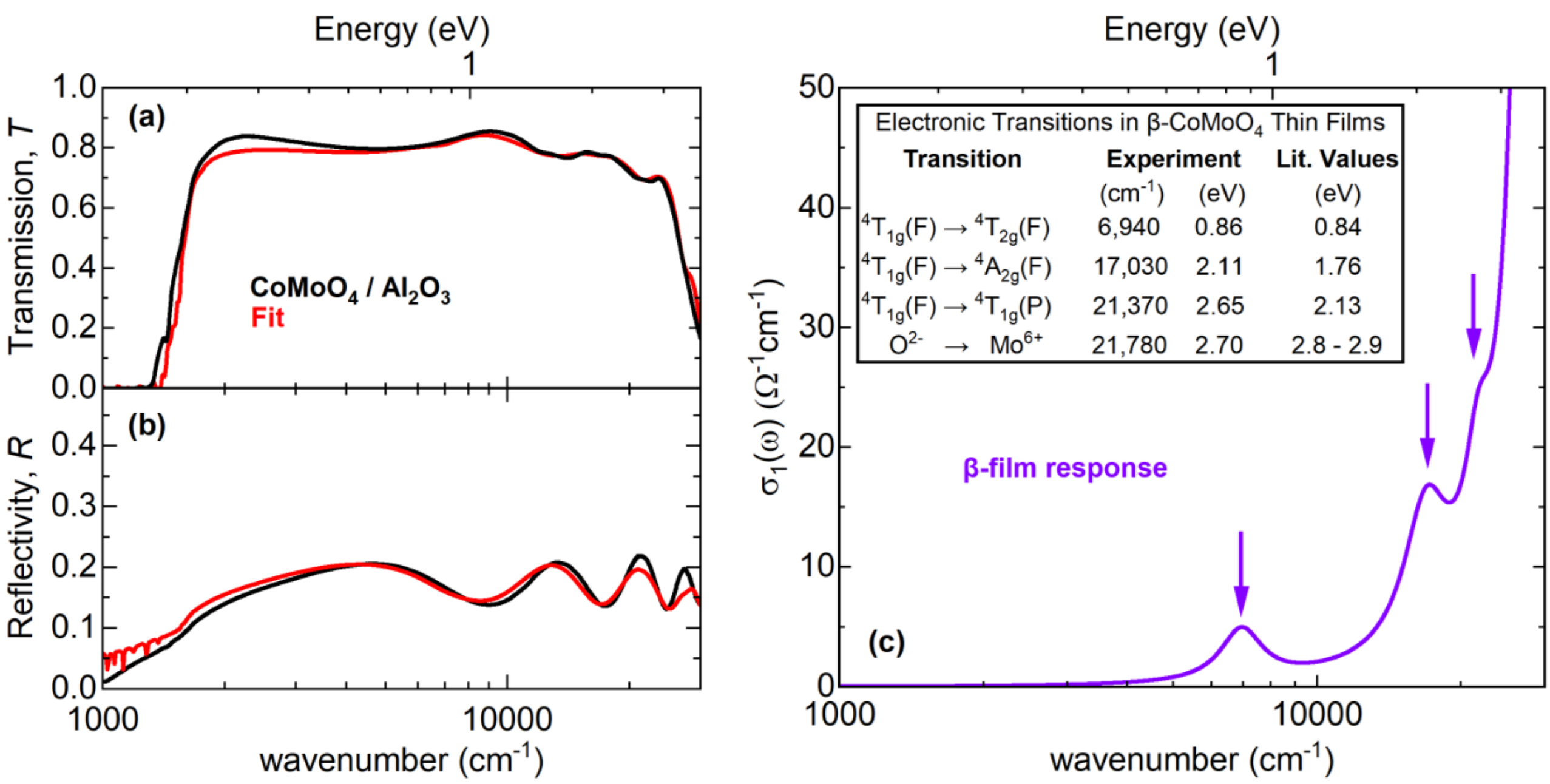

**Figure 6. Electronic Transitions – Experimental Results.** Experimental (black) and model (red) spectrum of the high-energy transmission **(a)** and reflectivity **(b)** for a 290 nm β-phase $CoMoO_4$ film grown on sapphire. **(c)** Derived optical conductivity $\sigma_1$ of the β-phase $CoMoO_4$ film response. The derived *d-d* and inter-band transitions for the β-phase $CoMoO_4$ film are highlighted by the arrows, and listed in comparison to the literature values, from Ref. [1,3,5,25]. As noted by Sreedhar et al. [47], the values are dependent on the crystal field environment, which is impacted by local strain.

There exist a few band structure calculations of the transition metal molybdates in the literature; however, in analogy to $ZnMoO_4$ [45] and $MnMoO_4$ [46], the valence band is dominated by oxygen *p*-orbitals and the conduction band by the Mo *d*-orbitals, while the 3*d* $Co^{2+}$ orbitals are largely located in the fundamental gap and near the conduction band edge. On the lower energy side, we can identify three further narrow absorption bands (highlighted by arrows in Figure 6(c)), which are assigned to the allowed *d-d* band transitions for $Co^{2+}$ in the $^4T_{1g}(F)$ ground state to excited states under the octahedral crystal field of $^4T_{2g}(F)$, $^4A_{2g}(F)$, and $^4T_{1g}(P)$ in accordance to Ref. [3]. The $Co^{2+}$ *d-d* band transition energies, reported in the inset of Figure 6(c), are derived from the Lorentzian fitting model that is used to describe the optical spectra in Figure 6(c), and are compared with the literature values for β-$CoMoO_4$. The agreement in comparison to powder samples is reasonable, by taking the likely high strain state of our films into account, which will alter the crystal field of the $Co^{2+}$ ions. This is further discussed in the following sections.

## Modified Crystal Field Theory

To elucidate the impact of strain on the *d-d* spectra, we calculate the changes in $Co^{2+}$ energy levels as the crystal field strength varies using a modified crystal field theory (MCFT) [48,49]. The result is shown in Figure 7. In the MCFT approach, the crystal field strength is parameterized by the effective nuclear charge of the metal ion- $Z_{eff}$, which decreases under the influence of the ligand's electron clouds compared to its value for the free ion. Thus, a decrease in the effective nuclear charge describes an increase in the crystal field strength. The MCFT input parameters are the valence state, the exact positions of the ligands in the crystal unit cell, as well as the position of the cation. The spin-orbit coupling is also accounted for [48]. We used the crystallographic data from the supplementary materials of Robertson et al. [3] and the data for the lattice constants of our β-phase thin film at 12 K from Table S1.

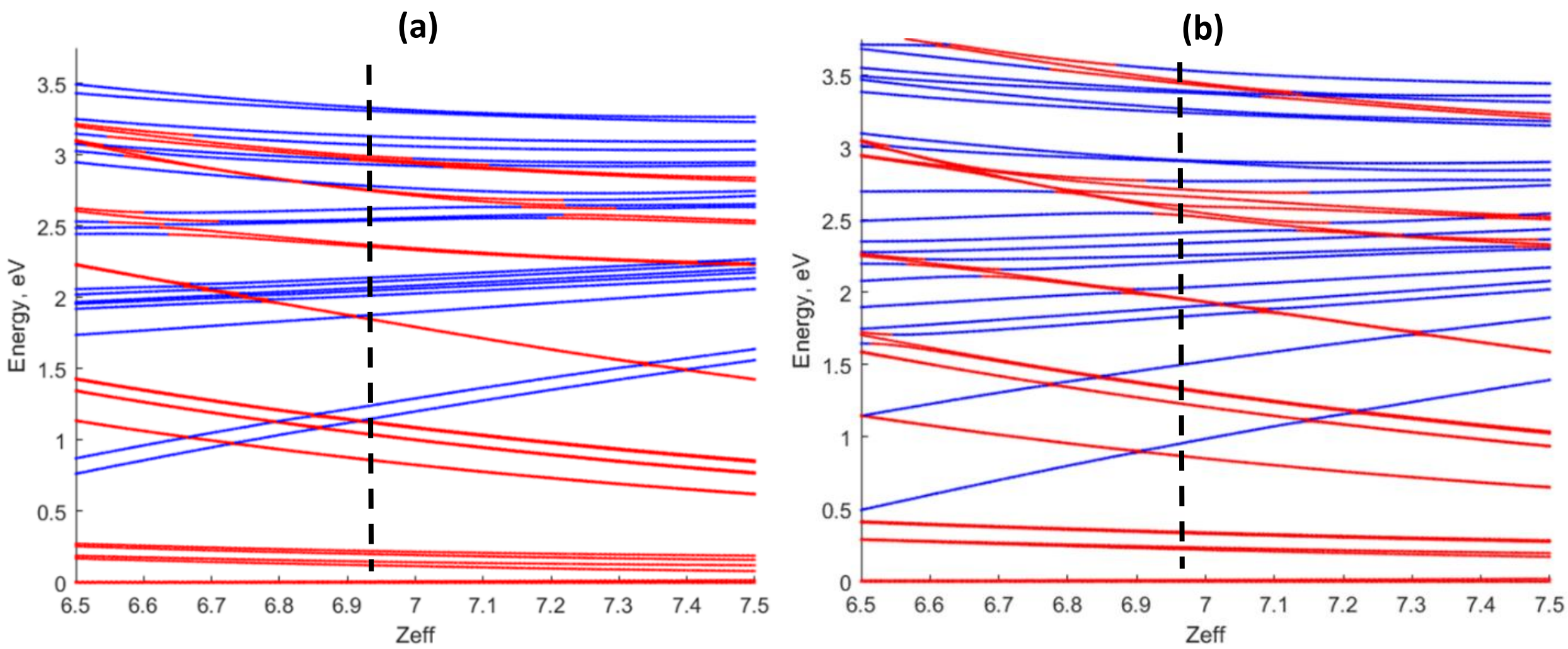

**Figure 7. Calculated energy diagrams**. The relationship between crystal field strength and energies of the $Co^{2+}$ $3d^7$ excitations are shown for **(a)** the $Co^{2+}$ ion in the 4i Wyckoff position ($C_s$-site symmetry), and **(b)** the $Co^{2+}$ ion in the 4h Wyckoff position ($C_2$-site symmetry). Here, the parameter $Z_{eff}$, presented in units of electron charge |e|, decreases with increasing crystal field strength. The calculated energies of the levels belonging to the $^4F$ and $^4P$ terms show a gradual increase with increasing crystal field strength. Vertical dashed lines denote $Z_{eff}$ = 6.925 in (a), and $Z_{eff}$ = 6.975 in (b) at which the theoretical energy levels well-coincide with experimentally observed bands at 0.86, 2.11 and 2.65 eV.

The energy diagrams for both positions of $Co^{2+}$ ions in the β-phase of $CoMoO_4$ film are shown in Figure 7. We demonstrate 38 Kramers doublets out of 60 of the $Co^{2+}$ $3d^7$ configuration, with energy below 3.75 eV in the range $6.5|e| < Z_{eff} < 7.5|e|$, with e being the electron charge. Note, for the free ion $Co^{2+}$ Zeff=8|e|, as reported in Ref. [50]. The energy levels originating from the $^4F$ and $^4P$ terms of the free ion (i.e. from the terms with the highest multiplicity 2S+1=4) are shown by red colored lines. It is expected that the most pronounced transitions, so-called "spin-allowed transitions", occur between the levels with the same multiplicity. The low site symmetry $C_2$ and $C_s$ at both $Co^{2+}$ positions results in a complete splitting of the levels initially assigned to a purely octahedral ligand environment. Note, that our calculations also suggest a splitting of the observed low-energy softmode centred at 40 $cm^{-1}$ when being exposed to an external magnetic field, which remains subject of future research.

As illustrated in the calculated energy diagrams in Fig. 7, the energy of the spin-allowed *d-d* transitions always show an increase with increasing crystal field strength. In our case, there are two almost degenerated Kramers doublets with energies in between 0.5 – 1 eV where a spin-allowed d-d transition may occur, i.e. the calculated energies of the two $Co^{2+}$ sites are insufficiently separated in energy to clearly assign the measured absorption bands to individual cobalt ions in the two different lattice positions. The energy of these *d-d* transitions can nonetheless serve as a characteristic marker of the crystal field strength in the $(CoO_6)$ cages. The experimentally obtained values for the hardening of the *d-d* band energies in our films are in good agreement relative to literature values as shown in the inset of Figure 6(c), which supports the evidence of a strained state of β-phase in our $CoMoO_4$ films. Note that the results of the MCFT calculations can be considered as a suitable guide for a qualitative explanation of the electronic spectra of Co ions under a change of the crystal field strength in the $CoMoO_4$ crystal. However, they can also give quantitative answer in the energy range far away from the ligands *p*-orbital energies, i.e. below the energy of the charge transfer transition band. In our case, this range is lower than 3 eV. Moreover, our method for analysing the thin films is different from that used on diffuse reflectance measurements of powders where the narrow absorption bands are located on the rising inter-band absorption region, which is an issue that becomes more prominent at higher energies where the discrepancy is at maximum.

## Summary

Thin films of β-$CoMoO_4$ were successfully synthesised on silicon and sapphire substrates, and their structural, magnetic, and optical properties were investigated. Temperature-dependent X-ray diffraction down to 12 K confirms that the films do not undergo the β→$\alpha$ transition observed in powders. Although the respective roles of particle size and microstrain cannot be fully disentangled, the data indicate that strain in the films governs the suppression of this thermochromic phase transition. Magnetic measurements down to 2.2 K confirm an antiferromagnetic transition near 10 K, consistent with our previous reports on powder samples, but with no magnetic hysteresis observed in the thin films. A field-induced transition at ±2.2 T is seen below the Néel temperature, with a sharp magnetisation increase that is characteristic of a spin-flip transition, indicating strong magneto-crystalline anisotropy; we argue that due to the sharpness of this transition it is not appropriate to classify it as a spin-flop transition, as observed in other molybdates. Room-temperature optical spectroscopy (terahertz to visible/UV) allows the identification of interband transitions and three $Co^{2+}$ *d-d* excitations, in good agreement with the literature. We also identify 16 infrared-active phonons. Upon cooling down to 150 K, being across the 1$^{st}$ order transition of β to $\alpha$-phase in powder samples, we observe that the higher-energy phonon modes (178 - 680 cm$^{-1}$) exhibit a hardening which is consistent with lattice contraction, while the lowest-energy mode at 42 cm$^{-1}$ softens, dominated by cation motion. This anomalous behaviour is attributed to incipient motion of $Mo^{6+}$ ions from their tetrahedral position in the β-phase to an octahedral site in the $\alpha$-phase, that is suppressed in thin film due to microstrain. Modified crystal field calculations support a strained state of the films and suggest a splitting of this mode when exposed to an applied magnetic field, which remains a subject of future research. These results demonstrate that strain stabilises the β-phase and modifies both lattice dynamics and electronic structure, providing a route to control functional properties in thermochromic oxides and relevant for the design of optical sensing devices.

## Acknowledgements

This work was financially supported by the New Zealand Synchrotron Group and the Australian Government ANSTO through proposal ID: 21184, Reference No: AS241/THz/21184 and the Marsden Fund of New Zealand (VUW1910). F.L. acknowledges support from the Swiss National Science Foundation through an Early Postdoc Mobility Fellowship with Project No. P2FRP2-199598, and the New Zealand Ministry of Business, Innovation & Employment through grant UOAX1710. The authors would like to acknowledge Christian Bernhard for helpful discussions, Blake Stanley for the support with some of the optical measurements, and the Surface Analysis Laboratory of Mark Wainwright Analytical Centre, The University of New South Wales, Australia, for the XPS analysis.